\documentclass[prl,twocolumn,superscriptaddress,amsmath,amssymb]{revtex4-1}
\usepackage{graphicx}
\usepackage{dcolumn,tabularx}% Align table columns on decimal point
\usepackage{bm}% bold math
\usepackage{braket}

\usepackage{mathdots}
\usepackage[T1]{fontenc}
\usepackage[unicode=true]{hyperref}
\usepackage{booktabs}
\usepackage{mathtools}

\usepackage{color}
\usepackage{soul}

\begin{document}
%\linenumbers

%\title{Remote entanglement distribution in a photonic quantum network via multinode indistinguishability}

\title{Remote entanglement distribution in a quantum network via multinode indistinguishability of photons}

\author{Yan Wang}
\affiliation{CAS Key Laboratory of Quantum Information, University of Science and Technology of China, Hefei 230026, China}
\affiliation{CAS Center for Excellence in Quantum Information and Quantum Physics, University of Science and Technology of China, Hefei 230026, China}
\affiliation{Hefei National Laboratory, University of Science and Technology of China, Hefei 230088, China}

\author{Ze-Yan Hao}
\affiliation{CAS Key Laboratory of Quantum Information, University of Science and Technology of China, Hefei 230026, China}
\affiliation{CAS Center for Excellence in Quantum Information and Quantum Physics, University of Science and Technology of China, Hefei 230026, China}
\affiliation{Hefei National Laboratory, University of Science and Technology of China, Hefei 230088, China}

\author{Zheng-Hao Liu}
\affiliation{CAS Key Laboratory of Quantum Information, University of Science and Technology of China, Hefei 230026, China}
\affiliation{CAS Center for Excellence in Quantum Information and Quantum Physics, University of Science and Technology of China, Hefei 230026, China}
\affiliation{Hefei National Laboratory, University of Science and Technology of China, Hefei 230088, China}

\author{Kai Sun}
\email{ksun678@ustc.edu.cn}
\affiliation{CAS Key Laboratory of Quantum Information, University of Science and Technology of China, Hefei 230026, China}
\affiliation{CAS Center for Excellence in Quantum Information and Quantum Physics, University of Science and Technology of China, Hefei 230026, China}
\affiliation{Hefei National Laboratory, University of Science and Technology of China, Hefei 230088, China}

\author{Jin-Shi Xu}
%\email{jsxu@ustc.edu.cn}
\affiliation{CAS Key Laboratory of Quantum Information, University of Science and Technology of China, Hefei 230026, China}
\affiliation{CAS Center for Excellence in Quantum Information and Quantum Physics, University of Science and Technology of China, Hefei 230026, China}
\affiliation{Hefei National Laboratory, University of Science and Technology of China, Hefei 230088, China}

\author{Chuan-Feng~Li}
\email{cfli@ustc.edu.cn}
\affiliation{CAS Key Laboratory of Quantum Information, University of Science and Technology of China, Hefei 230026, China}
\affiliation{CAS Center for Excellence in Quantum Information and Quantum Physics, University of Science and Technology of China, Hefei 230026, China}
\affiliation{Hefei National Laboratory, University of Science and Technology of China, Hefei 230088, China}

\author{Guang-Can Guo}
\affiliation{CAS Key Laboratory of Quantum Information, University of Science and Technology of China, Hefei 230026, China}
\affiliation{CAS Center for Excellence in Quantum Information and Quantum Physics, University of Science and Technology of China, Hefei 230026, China}
\affiliation{Hefei National Laboratory, University of Science and Technology of China, Hefei 230088, China}

\author{Alessia Castellini}
\affiliation{Dipartimento di Fisica e Chimica - Emilio Segr\`e, Universit\`a di Palermo, via Archirafi 36, 90123 Palermo, Italy}

\author{Bruno Bellomo}
\affiliation{Institut UTINAM, CNRS UMR 6213, Universit\'{e} Bourgogne Franche-Comt\'{e}, Observatoire des Sciences de l'Univers THETA, 41 bis avenue de l'Observatoire, F-25010 Besan\c{c}on, France}

\author{Giuseppe Compagno}
\affiliation{Dipartimento di Fisica e Chimica - Emilio Segr\`e, Universit\`a di Palermo, via Archirafi 36, 90123 Palermo, Italy}

\author{Rosario Lo Franco}
\email{rosario.lofranco@unipa.it}
\affiliation{Dipartimento di Ingegneria, Universit\`{a} di Palermo, Viale delle Scienze, Edificio 6, 90128 Palermo, Italy}

\begin{abstract}	

Quantum networking relies on entanglement distribution between distant nodes, typically realized by swapping procedures. However, entanglement swapping is a demanding task in practice, mainly because of limited effectiveness of entangled photon sources and Bell-state measurements necessary to realize the process. Here we experimentally activate a remote distribution of two-photon polarization entanglement superseding the need for initial entangled pairs and traditional Bell-state measurements. This alternative procedure is accomplished thanks to the controlled spatial indistinguishability of four independent photons in three separated nodes of the network, which enables us to perform localized product-state measurements in the central node acting as a trigger. This experiment proves that the inherent indistinguishability of identical particles supplies new standards for feasible quantum communication in multinode photonic quantum networks.

\end{abstract}

\maketitle

\section{Introduction}

Distributing quantum entangled states among nodes of a compound quantum network is the key process to implement large-scale and long-distance quantum information and communication processing \cite{Duan2001,Perseguers2008,PhysRevA.95.032306,Scherer:11,Das2018,Gyongyosi2019,Liu2021PRL,Lago2021,liu2021}. An established process to achieve this goal is provided by the so-called entanglement swapping, that entangles quantum particles which never mutually interact \cite{zukowski1993event,pan1998experimental,goebel2008multistage,ma2012experimental,zhang2019experimental,sun2017entanglement,Guccioneeaba4508,Basset2019}.
In this procedure the particles are spatially distinguishable and individually addressed by local operations and classical communication. Two initial entangled pairs of particles are thus required followed by Bell-state measurements (BSM) performed on two particles chosen each from a distinct pair, which finally leave the remaining two particles remotely entangled.

Implementation of entanglement swapping procedures, realized with different physical systems, such as photons \cite{pan1998experimental,RevModPhys.84.777,pan2002,yang2006experimental,halder2007}, atoms \cite{chou2007,monroe2008,pan2008}, diamond nitrogen-vacancy centers \cite{hanson2021}, and quantum dots \cite{SPDC2,basset2019,zopf2019}, has to face some drawbacks. On the one hand, entangled photon pair sources have typically limited efficiency, especially for those generated at low rate by spontaneous parametric down conversion (SPDC) \cite{zeilinger2009}. On the other hand, BSM are usually challenging to be realized with high performances \cite{BSA,cQED,cQED3,PhysRevA.84.042331,PhysRevA.92.042314,PhysRevA.82.032318}. One way to overcome these hurdles consists in obtaining technological improvements of the employed devices to increase generation and detection efficiency \cite{Jin2015}. Differently, one can think of a more fundamental route aiming at designing alternative methods which exploit and control inherent properties of the compound distributed system, such that their implementation is ultimately less demanding. In this work we follow the latter route, individuating in the spatial indistinguishability (SI) of identical particles the basic quantum property suited to the scope. %In this way, dual-rail entanglement is created over a larger distance is created only by means of single-photon sources.

Identical subsystems (e.g., photons, electrons, atoms of the same species) are typically the basic constituents of compound quantum systems \cite{braun2018quantum,yurke1984quantum,AltmanPRXQuantum,ladd2010quantum}. Moreover, proposals with identical particles serving as quantum resource in optical and cold atom systems have been provided \cite{lange2018entanglement,bellomoPRA2017,fadel2018spatial,kunkel2018spatially,
	killoran2014extracting,Blasiak2019,castelliniPhase,morris2020entanglement,lofranco2018ind}.
An operational framework to measure spatially indistinguishable (individually unaddressable) particles and generate entanglement is provided by spatially localized operations and classical communications (sLOCC) \cite{lofranco2018ind}. The sLOCC framework has been recently utilized experimentally with photonic setups \cite{sun2020experimental,Barros:20}, to realize teleportation \cite{sun2020experimental}, phase discrimination \cite{sunPNAS} and direct measurement of the exchange phase \cite{WangEP}, avoiding a physical particle exchange \cite{armandoEP,lofrancoNatPhot}. It is thus natural to proceed along this path for realizing entanglement distribution processes alternative to entanglement swapping.
Seeing the main issues of the standard entanglement swapping protocol mentioned above, it would be especially desirable to avoid the need for initial entangled pairs and possibly simplify the type of required final measurements. As suggested by recent theoretical studies \cite{castellini2019activating}, such a change to experimental entanglement distribution processes can be pursued by directly utilizing the intrinsic SI of identical particles which leads the particles in the intermediate node are not distinguishable.
Photons are particularly promising to this purpose due to the possibility to control the spreading of their wave packets towards separated sites, which in turn rules the degree of their spatial overlap at the nodes of interest. Based on this property, a method for achieving entanglement only by means of adjusting the SI between single-photons has been proved to be feasible \cite{sun2020experimental}.

Here we experimentally demonstrate heralded distribution of two-photon polarization entanglement between two remote nodes in a three-node quantum network without requiring initial entangled pairs and BSM. We employ four independent polarized photons whose wave packets are sent to three spatially separated nodes of the network and then measured by sLOCC. Suitably engineering the SI of photons in the nodes, we can perform localized product-state measurements (LPSM) in the central node which act as a polarization entanglement catalyst between the remote (extreme) nodes. Indeed, this is achieved just because we are running the experiment with indistinguishable particles. Besides its conceptual novelty, the key advantage of this process is that it simplifies the task of distributing entanglement, superseding the drawbacks encountered in the standard entanglement swapping with distinguishable particles during the initial preparation and the final measurement stages. The procedure, involving just single-photon sources, thus supplies advance in multinode quantum information and communication protocols.

\begin{figure}[t]
	\centering
	\includegraphics[width=0.48\textwidth]{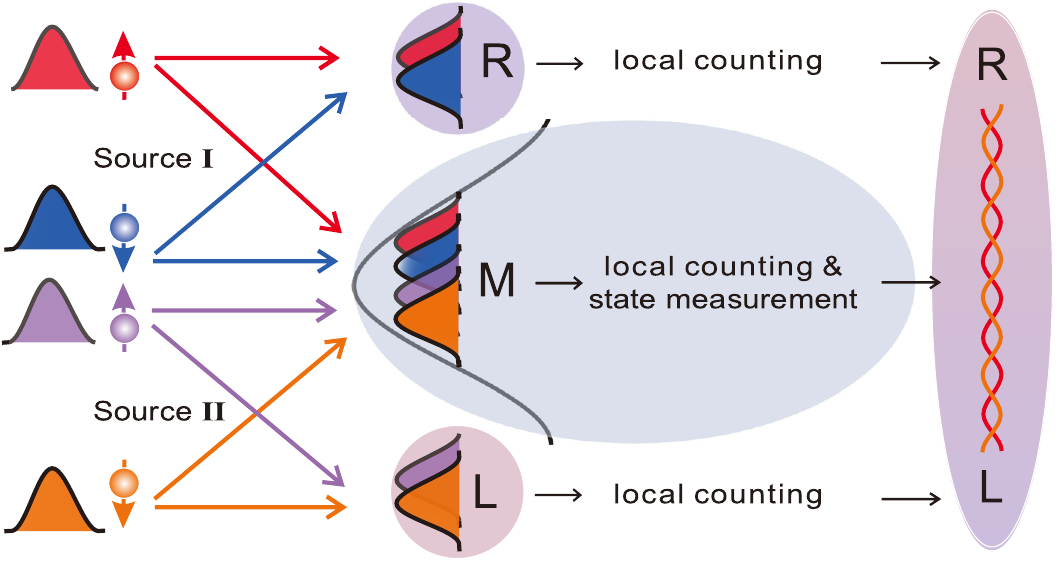}
	\caption{\textbf{Theoretical scheme.} Sources I and II generate two pairs of independent photons with opposite polarizations $\uparrow$ and $\downarrow$ which are sent to the spatially separated nodes R, M and L. Spatial modes of the particles coming from Source I (II) are adjusted to make them overlapping in nodes R, M (M, L). By performing local counting in each node and state measurements in the central node M, the photons in the remote nodes R and L get entangled.
	}
	\label{theory}
\end{figure}

\section{Theoretical background}

We consider a quantum network with three nodes, named R, M and L, and four initially independent (uncorrelated) photons. The aim is to distribute polarization entanglement between nodes R and L through an entanglement swapping-like procedure activated by controlling SI of the photons in the three nodes and sLOCC measurements.

Let us take two independent-photon sources, denoted as Source I and Source II in Fig.~\ref{theory}. Each source emits two uncorrelated photons with opposite pseudospins (e.g., polarizations) $\uparrow$ and $\downarrow$. One photon from Source I, with pseudospin $\uparrow$, is sent towards two spatially separated nodes R and M, giving the spatial mode $\ket{\alpha}=a\ket{\rm R}+b\ket{\rm M}$ ($|a|^2+|b|^2=1$): the spatial wave function can thus be written as a coherent superposition on the basis of nodes R and M. The other photon from Source I, with pseudospin $\downarrow$, is prepared in the spatial mode $\ket{\alpha'}=a'\ket{\rm R}+b'\ket{\rm M}$ ($|a'|^2+|b'|^2=1$).
Analogously, particles from Source II are prepared in
$\ket{\beta}=c\ket{\mathrm{M}}+d\ket{\mathrm{L}}$ with pseudospin $\uparrow$, and $\ket{\beta'}=c'\ket{\mathrm{M}}+d'\ket{\mathrm{L}}$ with pseudospin $\downarrow$ ($|c|^2+|d|^2=|c'|^2+|d'|^2=1$). Notice that $\ket{\alpha}$, $\ket{\alpha'}$, $\ket{\beta}$ and $\ket{\beta'}$ overlap in the central node M in which the particles cannot be distinguished in the spatial degree of freedom. As a result, the initial four-photon (unnormalized) state, expressed in the no-label formalism \cite{lofranco2016SciRep,compagno2018dealing} (see Appendix A for details), is $\ket{\Psi^{(4)}}=\ket{\alpha \uparrow, \alpha' \downarrow, \beta \uparrow, \beta' \downarrow}$.
The spatial overlaps in separated nodes are maximized to optimally exploit the effects of SI. Via sLOCC, locally counting two photons in the shared central node M and one photon in R and L (including classical communication of the outcomes), one obtains with a predicted success probability of $6/25$ (see Appendix B for details), the post-selected state \cite{castellini2019activating}
\begin{equation}
\label{PS}
\ket{\Psi^{(4)}_\mathrm{PS}}=\left(\ket{\Psi_\mathrm{M},\Psi^{+}_\mathrm{RL}}+\ket{\Phi^+_\mathrm{M},\Phi^+_\mathrm{RL}}-\ket{\Phi^-_\mathrm{M},\Phi^-_\mathrm{RL}}\right)/{\sqrt{3}},
\end{equation}
where $\ket{\Psi_\mathrm{M}}=\ket{\mathrm{M}\uparrow,\mathrm{M}\downarrow}$ and $\ket{\Phi^\pm_\mathrm{M}}=(\ket{\mathrm{M}\downarrow,\mathrm{M}\downarrow}\pm\ket{\mathrm{M}\uparrow,M\uparrow})/\sqrt{2}$ are Bell states in the M-subspace associated respectively to Bell states shared between R and L, namely $\ket{ \Psi_\mathrm{RL}^+}=(\ket{\mathrm{R}\uparrow, \mathrm{L}\downarrow}+\ket{ \mathrm{R}\downarrow, \mathrm{L}\uparrow})/\sqrt2$ and $\ket{ \Phi_\mathrm{RL}^\pm}=(\ket{ \mathrm{R}\uparrow,\mathrm{L}\uparrow}\pm\ket{ \mathrm{R}\downarrow,\mathrm{L}\downarrow})/ \sqrt2$.
Thus, two-photon polarization entanglement between remote nodes R and L is finally activated by distinguishing states $\ket{\Psi_\mathrm{M}}$, $\ket{\Phi^+_\mathrm{M}}$ (each occurring with probability $1/3$) on site M.
Furthermore, replacing photons with fermionic particles, the post-selected global state in Eq.~\eqref{PS} would become $\ket{\Psi_\mathrm{PS,f}^{(4)}}=\ket{\Psi_\mathrm{M},\Psi_\mathrm{RL}^-}$ ($\ket{ \Psi_\mathrm{RL}^-}=(\ket{\mathrm{R}\uparrow, \mathrm{L}\downarrow}-\ket{ \mathrm{R}\downarrow, \mathrm{L}\uparrow})/\sqrt2$) with a success probability of 2/9 \cite{castellini2019activating}, where states $\ket{\Phi^\pm_\mathrm{M}}$ vanish due to Pauli exclusion principle. We point out that here the entangled state between R and L is obtained with certainty after postselection, with no further measurements required.
Note that here, the entangled state between R and L could be determinately obtained.

Notice that these nodes remain unconnected during all the process and their entanglement catalyst is a localized state measurement in the central node M. This crucial role of the node M is made possible by SI of identical particles, inherently establishing accessible quantum correlations between separated couples of independent photons with no need of initial entangled pairs \cite{lofranco2018ind,sun2020experimental}. Also, we shall prove below that one can discriminate among the three M-subspace states by simple product-state measurements, so eluding BSM.

\begin{figure}[t]
	\includegraphics[width=0.49\textwidth]{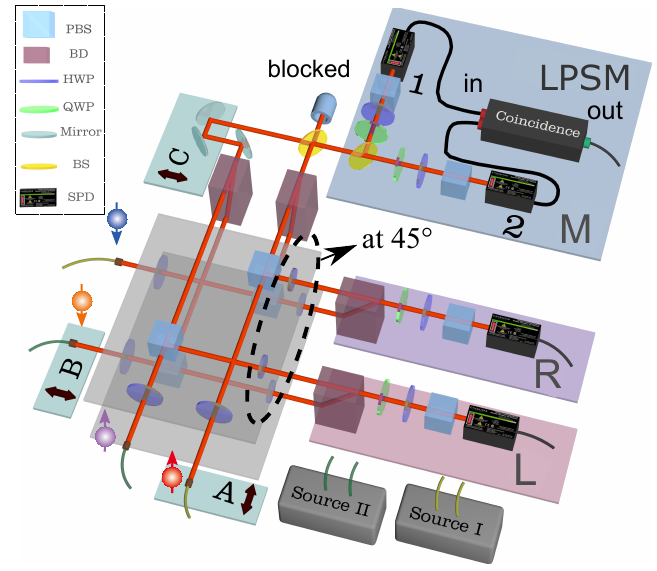}
	\caption{\textbf{Experimental setup.} Two photons from the same Source I (or II) with opposite spins respectively engineered and sent to nodes R and M (or M and L). Each of nodes R and L consists of one beam displacer (BD) and a polarization analysis unit made of a quarter-wave plate (QWP), a half-wave plate (HWP), and a polarized beam splitter (PBS). Two photons before node M pass the beam splitter (BS) whose one output is blocked by a shutter and the other output is manipulated by a second BS and polarization analysis unit. Then, the localized product-state measurement (LPSM) is performed in node M by projecting two arms (1 and 2) on different product states to activate entanglement distribution between R and L nodes with coincidence count of the signals from single-photon detectors (SPDs). And parts A, B and C are three movable platforms for adjustments of delay.
 }\label{experiment}
\end{figure}

\section{Experiment}
The setup of our experiment is displayed in Fig.~\ref{experiment}.
The pulsed ultraviolet beam (central wavelength at 400nm) consecutively passes through two beamlike type-II BBO crystals, delivering two pairs of photons with independent spatial wave functions via SPDC \cite{takeuchi2001beamlike}, respectively noted as Source I and Source II.  Both sources produce two independent photons with opposite polarization $\vert H\rangle$ (horizontal, $\ket{\uparrow}$) and $\ket{V}$ (vertical, $\ket{\downarrow}$) in the initially product states $\vert H\rangle\otimes\vert V\rangle$ (see Appendix C for experimental verification). The four product photons are then collected by single-mode fibers.
In principle, four independent single-photon sources could be used to start the process.
Here, the two BBO crystals are conveniently employed to generate independent product photons not requiring entangled pairs.

The four independent photons first pass the group of a half-wave plate (HWP), a polarized beam splitter (PBS), and another HWP at $45^\circ$ to ensure unvaried polarization (see dashed ellipse in Fig.~\ref{experiment}). Each group independently distributes the photon to remote nodes R, M (Source I), and M, L (Source II). Two photons from the same Source have a 3 mm vertical deviation and meet together after a beam displacer (BD) in the detection node (i.e., spatial overlap at every node). Polarization analysis units (PAUs) perform spatial local operations (sLOs) at corresponding nodes, while a coincidence device (CD) is used to handle the classical communication (CC). In addition, an interference filter (not shown) with the full width at half maximum of 3 nm is placed before each single-photon detector (SPD).
This all-optical setup has been designed to be as complicated as required to analyze the underlying physics, but not more.

\begin{figure*}[t]
	\centering
	\includegraphics[width=0.95\textwidth]{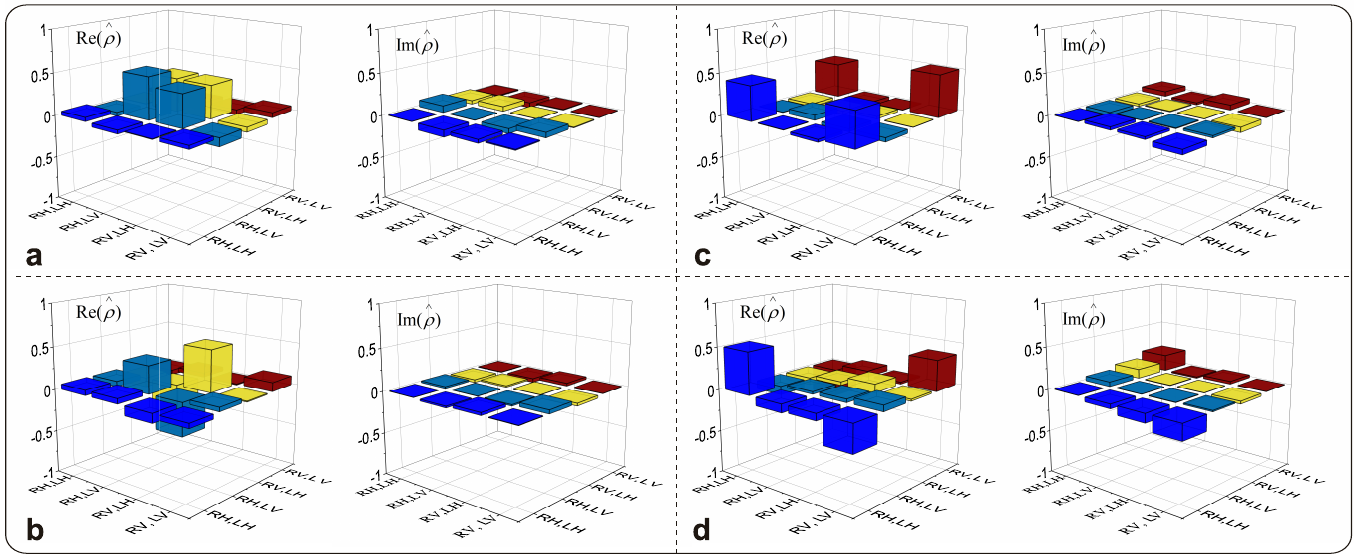}
	\caption{\textbf{Heralded remote entanglement.} Real and imaginary part of the reconstructed density matrix $\rho$ between nodes R and L after remote entanglement distribution. The predicted states between R and L are, respectively, $\ket{\Psi_\mathrm{RL}^{\pm}}=(\vert \mathrm{R}H, \mathrm{L}V\rangle\pm\vert \mathrm{R}V,\mathrm{L}H\rangle)/\sqrt2$ in panels \textbf{a, b} and $\ket{\Phi_\mathrm{RL}^{\pm}}=(\vert \mathrm{R}H,\mathrm{L}H\rangle \pm\vert \mathrm{R}V,\mathrm{L}V\rangle)/\sqrt2$ in panels \textbf{c, d}.}\label{swapping_fig}
\end{figure*}

Firstly, the Hong-Ou-Mandel (HOM) interference \cite{hong1987measurement} is performed to characterize the degree of temporal indistinguishability among photons. Results show that photons from two sources are generated with visibilities of 0.9734 $\pm$ 0.0032 and 0.9593 $\pm$ 0.0045, respectively. Indistinguishability among three nodes is observed with the interference dip of two photons overlapped via a beam splitter in node M while the other sides (nodes R and L) are treated as triggers, with a visibility of $0.8436\pm0.0405$. All the value of the above visibilities are calculated including the
limiting factors from accidental coincidence counts and multi-photon excitation, see Appendix C for more details. These procedures ensure the four independent photons of network are all indistinguishable.

The desired photon state from Sources I and II is prepared as $\ket{\alpha H,\alpha^\prime V,\beta H,\beta^\prime V}$. By tuning the angle of the corresponding HWP before the input PBS, we can change the probability amplitudes of the photons.
To maximize the SI of the photons in nodes R, M, and M, L, we set $\ket{\alpha}=\ket{\alpha'}=(\vert \mathrm{R}\rangle+\vert \mathrm{M}\rangle)/\sqrt{2}$, $\ket{\beta}=\ket{\beta'}=(\vert \mathrm{M}\rangle+\vert \mathrm{L}\rangle)/\sqrt2$. This means a photon is sent to nodes R, M or L with balanced probability, allowing maximal success probability of entanglement activation between R and L (find more details in Appendix D).

The theoretical analysis demands only one spatial mode in the central node M to implement the local counting. To satisfy this requirement, the two paths before node M are combined together via the first beam splitter with one side blocked and the other sent for product measurement, as shown in Fig.~\ref{experiment}, allowing us to exploit the spatial indistinguishability of photon particles. This local counting procedure requires to distinguish all three possible two-qubit states in Eq.~\ref{PS}, which is fulfilled by performing LPSM in node M to deal with this one-node case. However, inherently different from LPSM, the BSM, which is usually used to realize the standard entanglement swapping with projecting the outputs on the Bell states between two intermediate nodes, handles the two-mode case in which there are two separated spatial modes before the measurement stage \cite{pan1998experimental,RevModPhys.84.777}.
It is also worth to mention that the hyperentanglement involving multiple degrees of freedom is able to achieve complete and deterministic Bell-state measurement \cite{PhysRevLett.96.190501}.
Here, the LPSM is realized by splitting the one spatial mode into two outputs ``1'' and ``2'', as shown in Fig.~\ref{experiment}, and then projecting two outputs on a product basis to distinguish three states of Eq.~\ref{PS}
. Then we could activate the entanglement between the remote and unconnected nodes R and L via the LPSM.

The detailed process is introduced as follows. By placing another BS in the node M to split the beam into two paths ``1'' and ``2'', the state $\ket{\Psi_\mathrm{M}}$ becomes $(\ket{H_1, V_2}+\ket{V_1, H_2}-i\ket{H_1, V_1}+i\ket{H_2, V_2})/2$ considering the effect of the first BS. Because of the role of post-selection measurement, the state $\ket{\Psi_\mathrm{M}}$ effectively transforms into $\ket{\Psi'_\mathrm{M}}=(\ket{H_1, V_2}+\ket{V_1, H_2})/\sqrt2$. Similarly, states $\ket{\Phi_\mathrm{M}^{\pm}}$ would turn into $\ket{\Phi_\mathrm{M}'^{\pm}}=(\ket{H_1, H_2}\pm\ket{V_1, V_2})/\sqrt2$ (see Appendix E for details).
The LPSM enables to distinguish three states $\ket{\Psi'_\mathrm{M}}$ and $\ket{\Phi_\mathrm{M}'^{\pm}}$ by setting appropriate product measurements on the two outputs of M node, where two of these three states vanish while the other one is nonzero, e.g., measuring the product state $\ket{H_1} \otimes \ket{V_2}$ (or $\ket{V_1} \otimes \ket{ H_2}$) in node M, the outputs of $\ket{\Phi_\mathrm{M}'^{\mp}}$ are zero while outcomes of $\ket{\Psi'_\mathrm{M}}$ would exhibit coincidence at two final detectors in M.
% In particular: (i) measuring the product state $\ket{H_1} \otimes \ket{V_2}$ (or $\ket{V_1} \otimes \ket{ H_2}$) on node M, the outputs of $\ket{\Phi_\mathrm{M}'^{\mp}}$ are zero while the outcomes of $\ket{\Psi'_\mathrm{M}}$ would exhibit coincidence at the two final detectors in M; (ii) projecting on $\ket{r_1} \otimes \ket{l_2}$ (or $\ket{l_1} \otimes \ket{r_2}$), where $\ket{r}=(\ket{H}+i\ket{V})/\sqrt2$ and $\ket{l}=(\ket{H}-i\ket{V})/\sqrt2$, we pick up $\ket{\Phi_\mathrm{M}'^{+}}$ since $\ket{\Phi_\mathrm{M}'^{-}}$ and $\ket{\Psi'_\mathrm{M}}$ vanish; (iii) the state $\ket{\Phi_\mathrm{M}'^{-}}$ is filtered out by measuring the product state $\ket{d_1} \otimes \ket{c_2}$ (or $\ket{c_1} \otimes \ket{d_2}$), where $\ket{d}=(\ket{H}+\ket{V})/\sqrt2$ and $\ket{c}=(\ket{H}-\ket{V})/\sqrt2$.
Notably, the output signal of node M is the twofold coincidence count of detectors placed on paths ``1'' and ``2'' in LPSM, and then it is further used to coincide with outputs from L and R. Thus, the three-node coincidence from M, L and R (which is actually a fourfold coincidence) is used for tomography of entanglement distributed between nodes R and L.
Additionally, if the two photons from one source (I or II) both arrive at the node M where the three-node coincidence is not available, this leads to the failure of activation of entanglement between node R and L even if the output of M would still have count.
%Additionally, if the two photons from one source (I or II) both arrive at the node M, the output of M would still have counts but the threefold coincidence is not available, which leads to the failure of activation of entanglement between node R and L.

Placing node M on different bases, we separately pick $\ket{\Psi'_\mathrm{M}}$, $\ket{\Phi_\mathrm{M}'^{+}}$, and $\ket{\Phi_\mathrm{M}'^{-}}$.
Depending on the outcome, two remote and unconnected nodes R and L eventually share entangled states $\ket{\Psi_\mathrm{RL}^+}=(\ket{\mathrm{R}H,\mathrm{L}V}+\ket{\mathrm{R}V,\mathrm{L}H})/\sqrt2$ and $\ket{\Phi_\mathrm{RL}^{\pm}}=(\ket{\mathrm{R}H,\mathrm{L}H}\pm\ket{\mathrm{R}V,\mathrm{L}V})/\sqrt2$.

On the other hand, the only distributed entangled state $\ket{\Psi_\mathrm{RL}^-}$ reached for fermions, can be directly achieved in our setup by adding a relative phase $\pi$ on either R or L node provided that $\ket{\Psi_\mathrm{M}'}$ is measured in M. In the actual experiment, the additional phase $\pi$ is applied to node R (implying $\ket{\alpha'}= -a'\ket{\mathrm{R}}+b'\ket{\mathrm{M}}$). Node M is in the measurement basis $\{\ket{H},\ket{V}\}$ to pick $\ket{\Psi'_\mathrm{M}}$ and the obtained state $\ket{\Psi_\mathrm{RL}^-}$ is given by the experimental density matrix in Fig.~\ref{swapping_fig}{\bf b}. Alternatively, if no additional phase is added, $\ket{\Psi_\mathrm{RL}^-}$ can be produced by a local $\sigma_z$ operation on either R or L node after picking $\ket{\Psi_\mathrm{M}'}$.

\begin{table}[t!]
\caption{Experimental results of density matrices between nodes R and L after entanglement distribution, consisting of data of fidelity $F$ and concurrence $C$.
	}
	\begin{ruledtabular}
		\begin{tabular}{ccccc}
			state & $F$ & error-bar & $C$ & error-bar   \\ \hline
			$\vert \Psi^+_\mathrm{RL}\rangle$ & 0.8392 & 0.0381 & 0.7416 & 0.0737  \\
			$\vert \Psi^-_\mathrm{RL}\rangle$ & 0.8731 & 0.0228 & 0.7842  & 0.0473  \\ \hline
			$\vert \Phi^+_\mathrm{RL}\rangle$ & 0.8524 & 0.0231 & 0.7261 & 0.0766  \\
			$\vert \Phi^-_\mathrm{RL}\rangle$ & 0.8236 & 0.0371 & 0.7598  & 0.0903 \\
		\end{tabular}
	\end{ruledtabular}\label{table:1}
\end{table}

%Along the single-beam path at node M which outputs the coincidence count of paths ``1'' and ``2'', we detect the coincidence of three outputs and perform state tomography at the nodes R and L which includes single-photon counting, considering the central node M as the trigger
The experimental data of these outputs are given in Fig.~\ref{swapping_fig} and Tab.~\ref{table:1} (not subtracting accidental coincidence counts), reporting the fidelity $F=\rm{Tr}[\sqrt{\rho_{\mathrm{th}}}\rho_\mathrm{exp}\sqrt{\rho_{\mathrm{th}}}]$ ($\rho_{\mathrm{th}}$ and $\rho_\mathrm{exp}$ being, respectively, the theoretical state and the experimental state) and the concurrence $C$ \cite{nielsen2002quantum,PhysRevLett.80.2245} to quantify the actual entanglement \cite{horodecki2009quantum}. All the results are higher than the classical fidelity limit $2/3$ \cite{jennewein2001experimental}. Thus, because of photon SI, the sLOCC procedure here performed (local particle counting, product-state measurements and coincidence) allows one to generate nonlocal entangled states shared by remote nodes R and L.

\section{Discussion.}
We have reported the experimental realization of heralded entanglement distribution between separated distant nodes activated by controlling SI of four independent photons in a three-node quantum network. The setup has been suitably designed to efficiently realize the process via sLOCC at nodes R, M, and L. Specifically, we have implemented sLO and LPSM in a central node M and single-photon counting in nodes R and L, where photons have to be remotely entangled (together with coincidence measurements, i.e., CC), experimentally confirming previous theoretical predictions \cite{castellini2019activating}.

Differently from standard entanglement swapping protocols with distinguishable particles and LOCC, neither initial entangled photon pairs nor BSM are needed in our quantum network when realizing the remote entanglement distribution via sLOCC. The four initial photons are independent, uncorrelated and remain unconnected before arriving at the detection nodes. The polarization entanglement catalyst between photons in remote nodes R and L is the LPSM on two photons in the central node M. We remark that the key quantum property allowing such a process is the SI of the independent photon pairs in the network nodes which inherently leads to accessible quantum correlations (entanglement) \cite{lofranco2018ind,sun2020experimental}. The same process does not work if distinguishable (individually addressed) particles travel through the network, just giving separable states at the nodes. This method for achieving dual-rail entanglement \cite{RevModPhys.79.135} with just single-photon sources may possess significant potential in implementing more scalable quantum internet since the technology of on-demand single-photon sources are more reliable than on-demand entanglement sources.

Compared with previous works where single-photon sources are considered to implement entanglement distribution between two spatial modes based on Fock states with BSM \cite{gisin2007,chou2007}, here we have specifically engineered and performed the LPSM, thanks to the possibility of operating along the single-beam path in the central node M, superseding the BSM procedure. This feature emphasizes how localized operations in the sLOCC framework with indistinguishable particles are capable to activate entanglement among remote nodes, resulting even more advantageous. Despite the realized entanglement distribution process is conditional due to the success probability associated to the sLOCC, it overcomes the demanding requirements of initial entangled pairs (whose joint probability can be estimated around $10^{-4}$ \cite{SPDC2}) and complete BSM. These results, showing both conceptual and practical advance, open new avenues of linear optical networks for quantum information \cite{PhysRevLett.126.123601}. Also, recent theoretical findings about noise-protected entanglement by SI \cite{indistdynamicalprotection,indistentanglprotection,e23060708,piccoliniArXiv}, together with an experiment of indistinguishability-enabled coherence endurance \cite{armandoNPJ}, suggest that our entanglement distribution protocol can be made robust (to be studied elsewhere).

We highlight that the realized scheme is in principle applicable to fermions, provided that a platform implementing linear optics with electrons (or other fermions) is employed. For instance, one may think of platforms adopting quantum dots as sources of single electrons that can be initialized in particular spin states \cite{Yamamoto2008}, emitted on demand \cite{feve2007demand} and directed to quantum point contacts acting as electronic beam splitters \cite{bocquillon2013coherence,electronsReview}. This kind of experiments is left as a further route of investigation for future works. 

The four-photon three-node quantum network realized in this work constitutes the basic setup straightforwardly generalizable to multinode configurations. This experiment hence has the role of a proof-of-principle which may motivate the design of ameliorated protocols based on the same mechanisms. SI of photons thus promises to be further exploited in high-dimensional quantum communication scenarios \cite{morris2020entanglement,PhysRevLett.125.123603}, paving the way to new possible standards for large-scale quantum networks.

%\cite{ritter2012elementary,kalb2017entanglement}.

%This work emphasizes the generated entanglement built on the indistinguishable particles could act as the physical resource in the composite quantum system and scalable quantum information tasks, which is a must to achieve large-scale quantum information distribution tasks and a key process to quantum communication \cite{PhysRevA.95.032306,Scherer:11}.

\begin{acknowledgments}
This work was supported by the National Key Research and Development Program of China (Grant No. 2017YFA0304100), National Natural Science Foundation of China (Grants Nos. 11821404, 11774335, 61725504, 61805227, 61975195, U19A2075), Anhui Initiative in Quantum Information Technologies (Grants Nos. AHY060300 and AHY020100), Key Research Program of Frontier Science, CAS (Grant No. QYZDYSSW-SLH003), Science Foundation of the CAS (No. ZDRW-XH-2019-1), the Fundamental Research Funds for the Central Universities (Grant No. WK2470000026), Innovation Program for Quantum Science and Technology (No. 2021ZD0301200).
R.L.F. acknowledges support from European Union -- NextGenerationEU -- grant MUR D.M.
737/2021 -- research project ``IRISQ''.
\end{acknowledgments}

%\bibliography{ref1}% Produces the bibliography via BibTeX.
\setcounter{figure}{0}
\setcounter{equation}{0}
\renewcommand\thefigure{S\arabic{figure}}
\renewcommand\theequation{S\arabic{equation}}

\section*{appendix A:~No-label formalism and probability amplitudes}\label{app:A}

\newcommand{\scal}[2]{\langle#1|#2\rangle}

In the no-label approach, the global state is taken as the set of one-particle states and is to be considered as a whole entity, since the particles are in general unaddressable individually \cite{lofranco2016SciRep}.
Let us take the simple case of a system of two identical particles, whose state vector describes one particle in the state $\ket{\varphi_1}$ and one in $\ket{\varphi_2}$ (notice that there is no label associated to the particles; we do not know which particle has a given state). The global state is completely characterized by enumerating the single-particle states and represented as $\ket{\Phi^{(2)}}=\ket{\varphi_1,\varphi_2}$. The physical predictions on the system follow from the two-particle probability amplitudes defined as
\begin{equation}\label{scalarproduct}
	\scal{\varphi'_1,\varphi'_2}{\varphi_1,\varphi_2}:= \scal{\varphi'_1}{\varphi_1}\scal{\varphi'_2}{\varphi_2}+\eta\scal{\varphi'_1}{\varphi_2}\scal{\varphi'_2}{\varphi_1},
\end{equation}
where $\eta=\pm1$ for bosons and fermions, respectively. Notice that each single-particle state contains all the relevant degrees of freedom. For example, if the state $\ket{\varphi_i }$ ($i=1,2$) is characterized by a spatial degree of freedom (spatial wave function) $\psi_i$ and a pseudospin $\sigma_i$, then $\ket{\varphi_i}=\ket{\psi_i\sigma_i}$ and the two-particle state is written as $\ket{\Phi^{(2)}}=\ket{\psi_1\sigma_1,\psi_2\sigma_2}$. Since different degrees of freedom are independent, the probability amplitudes (scalar products) will clearly exhibit single-particle products of corresponding degrees of freedom, that is
\begin{equation}
	\begin{aligned}
		\label{scalarproduct2}
		\scal{\varphi'_1,\varphi'_2}{\varphi_1,\varphi_2}&=
		(\scal{\psi'_1}{\psi_1} \scal{\sigma'_1}{\sigma_1})(\scal{\psi'_2}{\psi_2}\scal{\sigma'_2}{\sigma_2})\nonumber\\
		&+\eta(\scal{\psi'_1}{\psi_2} \scal{\sigma'_1}{\sigma_2})(\scal{\psi'_2}{\psi_1}\scal{\sigma'_2}{\sigma_1}).
	\end{aligned}
\end{equation}

The above definitions can be straightforwardly generalized to a system of $n$ identical particles \cite{compagno2018dealing}. If the single-particle states are $\ket{\varphi_i}$ ($i=1,2,\ldots,n$), the global state is represented by listing the single-particle states, that is: $\ket{\Phi^{(n)}}=|\varphi_1,\varphi_2,\ldots,\varphi_n\rangle$.
For calculating the transition (success) probabilities under different configurations, we need to compute scalar products between states of $n$ identical particles. These are obtained by the $n$-particle probability amplitude defined as \cite{compagno2018dealing}
\begin{eqnarray}\label{Nampleta}
	&\langle \varphi'_1,\varphi'_2,\ldots,\varphi'_n|\varphi_1,\varphi_2,\ldots,\varphi_n\rangle & \nonumber\\ & :=\sum_P\eta^P\langle \varphi'_1|\varphi_{P_1}\rangle\langle \varphi'_2|\varphi_{P_2}\rangle \ldots \langle \varphi'_n|\varphi_{P_n}\rangle, &
\end{eqnarray}
where $P=\{P_1,P_2,...,P_n\}$ in the sum runs over all the one-particle state permutations, $\eta=\pm1$ for bosons and fermions, respectively, and $\eta^P$ is 1 for bosons and 1 (-1) for even (odd) permutations for fermions.
Notice that the probability amplitude above are also useful to determine the normalization constant of a global state of $n$ identical particles.

\section*{appendix B: Success probability after sLOCC post-selection}\label{app:B}
We give in the following the details about the calculations to determine the success probability after sLOCC to obtain the state $\ket{\Psi^{(4)}_\mathrm{PS}}$ of Eq.~(1) of the main text. The original calculations are reported in Ref.~\cite{castellini2019activating}. The state $\ket{\Psi^{(4)}_\mathrm{PS}}$ is the one effectively produced in our experiment.

The normalized initial four-boson global state $|\Psi_\mathrm{nor}^{(4)}\rangle=\dfrac{1}{\mathcal{N}}\ket{\alpha \uparrow, \alpha' \downarrow, \beta \uparrow, \beta' \downarrow}$ can be explicitly written as
\begin{equation}
	|\Psi_\mathrm{nor}^{(4)}\rangle=\dfrac{1}{5}|\mathrm{(R+M)\downarrow,(R+M)\uparrow,(M+L)\downarrow,(M+L)\uparrow\rangle}.
\end{equation}
The basis for sLOCC, corresponding to counting two particles in the shared intermediate node M and one particle in each mode R and L, is $\mathcal{B}=\Bigl\{\dfrac{|\mathrm{R \ \sigma, M \ \tau, M \ \sigma', L \ \tau'\rangle}}{\mathcal{N}_{\tau\sigma'}}\Bigr\}$ \ ($\sigma,\tau,\sigma',\tau'=\downarrow,\uparrow$) where $\mathcal{N}_{\tau\sigma'}=\sqrt{1+\langle \tau|\sigma'\rangle}$.
The four-boson post-selected state, after sLOCC projection onto $\mathcal{B}$, is then
\begin{align}\label{finalb}
	\begin{split}
		\ket{\Psi^{(4)}_\mathrm{PS}}=&\dfrac{1}{\sqrt{6}}(|\mathrm{R}\downarrow,\mathrm{M}\uparrow,\mathrm{M}\uparrow,\mathrm{L}\downarrow\rangle+|\mathrm{R}\downarrow,\mathrm{M}\uparrow,\mathrm{M}\downarrow,\mathrm{L}\uparrow\rangle\\
		&+|\mathrm{R}\uparrow,\mathrm{M}\uparrow,\mathrm{M}\downarrow,\mathrm{L}\downarrow\rangle+|\mathrm{R}\uparrow,\mathrm{M}\downarrow,\mathrm{M}\downarrow,\mathrm{L}\uparrow\rangle).
	\end{split}
\end{align}
The sLOCC (success) probability to obtain the state above is given by $\mathcal{P}=|\langle \Psi^{(4)}_\mathrm{PS}|\Psi_\mathrm{nor}^{(4)}\rangle|^2=6/25$, where we have used a sum of expressions of the kind of Eq.~(\ref{Nampleta}) with $\eta=1$ and $n=4$.

\section*{APPENDIX C: Sources of four indistinguishable photons}\label{app:C}

\begin{figure}[h]
	\centering
	\includegraphics[width=0.5\textwidth]{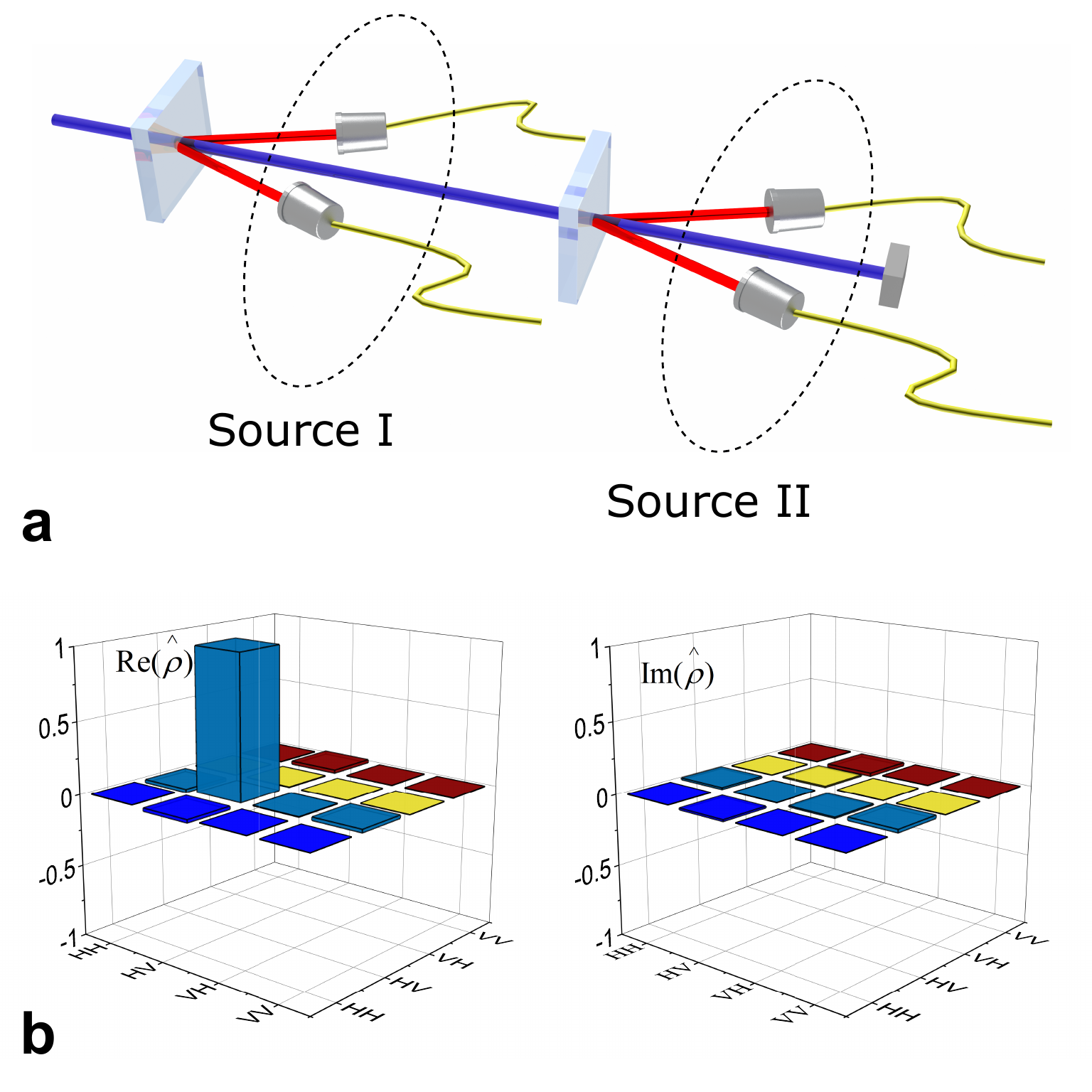}
	\caption{{\bf a.} The two independent photons from spontaneous parametric down conversion procedure. {\bf b.} The corresponding density matrix of $\ket{H}\otimes\ket{V}$ with fidelity of $99.9\pm0.01\%$ \cite{sun2020experimental}.} \label{source}
\end{figure}

Two photons emitted from a spontaneous parametric down conversion procedure (see Fig.~\ref{source}{\bf a}) share a polarized product state $\ket{H}\otimes\ket{V}$ in which there is no quantum correlations, as experimentally verified in Fig.~\ref{source}{\bf b}. Therefore, each source employed in the setup generates independent single photons.

\begin{figure}[h]
	\centering
	\includegraphics[width=0.48\textwidth]{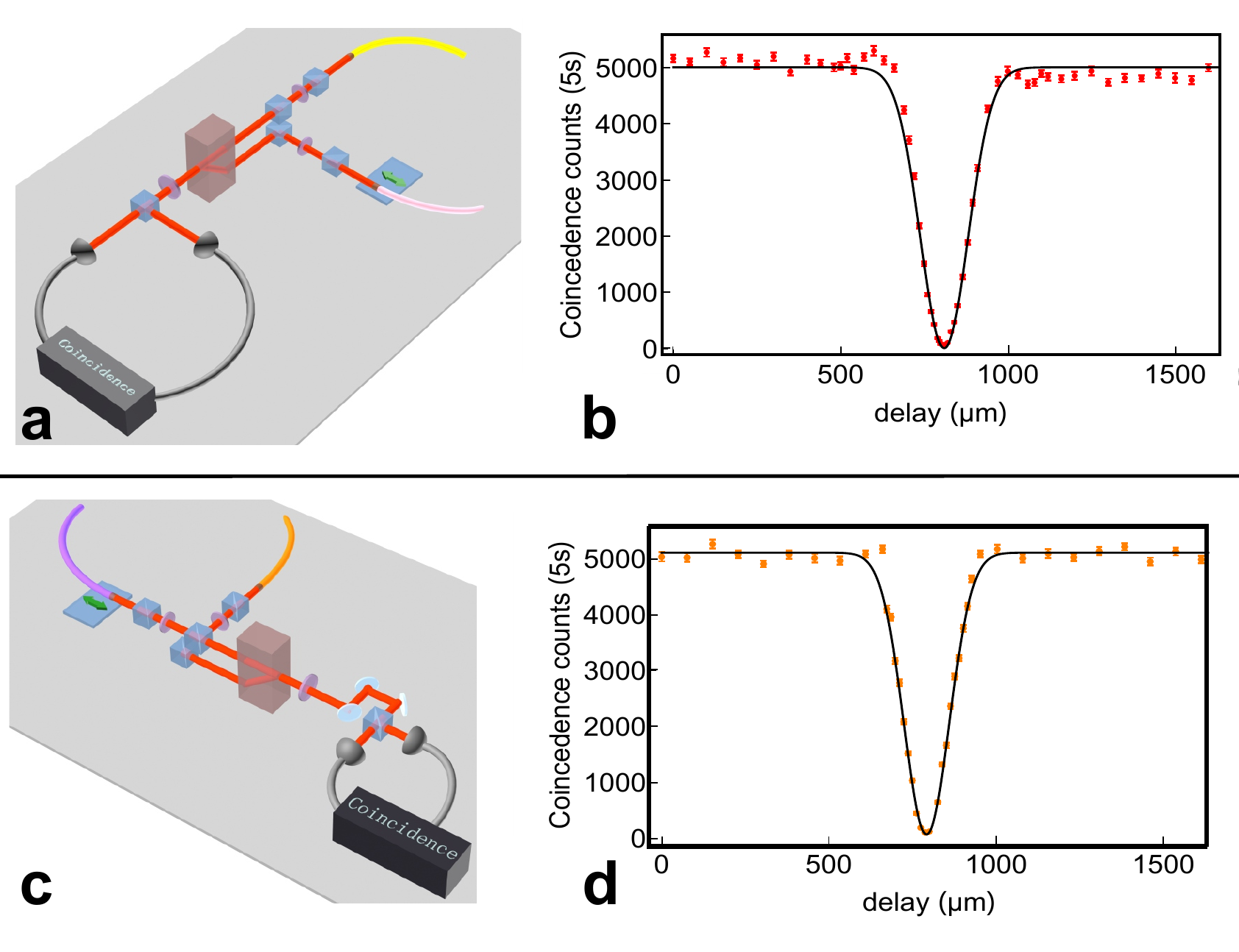}
	\caption{The experimental Hong-Ou-Mandel (HOM) interference setup and results of the photon pairs. \textbf{a.} HOM experimental setup of photons from Source I; \textbf{b.} The HOM dip corresponding to Source I, where the red points are the experimental dots and the black solid line is the theoretical curve. \textbf{c.} Interference setup of photons from Source II; \textbf{d.} The HOM dip of the photon pair from Source II, where the orange points are the experimental dots and the black solid line is the theoretical curve.} \label{HOM}
\end{figure}

The premise of our experiment is the indistinguishability among initially producted photons, which directly determines the quality of final entanglement shared by nodes R and L. Therefore, we should adjust the spatial overlap of the particles to make them all identical and coherent through HOM interference. As displayed in Fig.~\ref{HOM}, photons from the yellow and pink SMFs are emitted from Source I \textbf{a} and photons from purple and orange SMFs are from Source II \textbf{c}, where the other SMFs are depicted with different colors just for the sake of clarity. Taking the two photons with opposite spins from Source I as example, they separately pass the respective PBS, HWPs and PBSs with a vertical distance of 3 mm, and meet together at the detection node after the BDs. The followed HWP at $22.5^{\circ}$ is used to project the photons on $(\vert H\rangle+\vert V\rangle)/\sqrt2$. Finally the photons are collected at two parts after passing PBS for coincidence, where an interference filter (not shown) with the full width at half maximum of 3 nm is placed before collection.

\begin{figure}[h]
	\centering
	\includegraphics[width=0.48\textwidth]{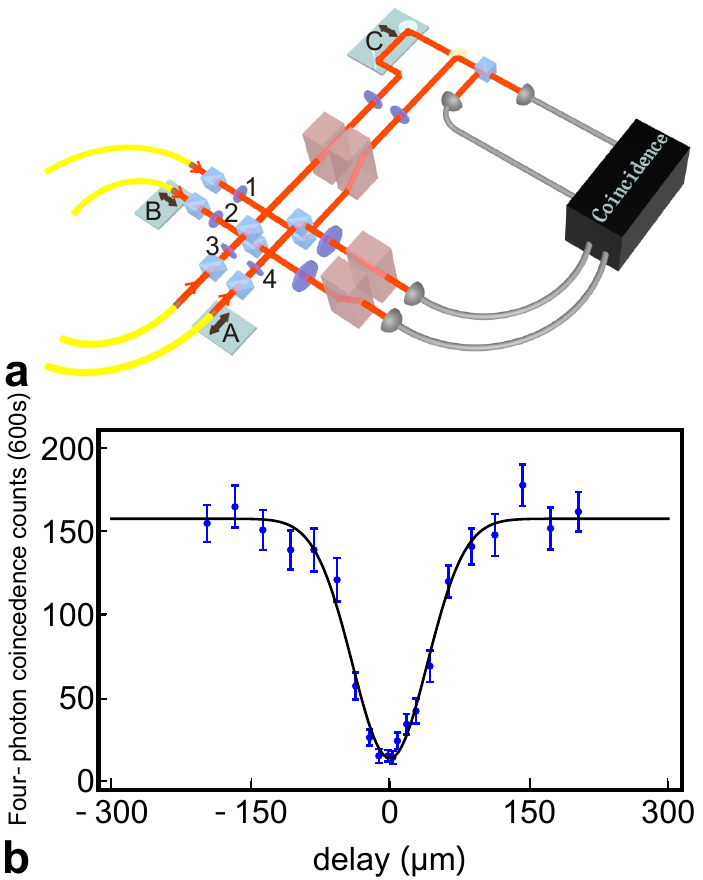}
	\caption{\textbf{a.} The experimental setup to detect the four-photon indistinguishability. Four photons respectively pass the PBS, HWP and PBS, and the following HWPs at $45^\circ$ for a fixed polarization. Differently from the main experiment illustrated in the manuscript, the beam displacers (BDs) are useless in this case. The HWPs 2 and 3 are placed at $0^\circ$, while HWPs 1 and 4 are at $45^\circ$ for a higher coincidence count rate. One side (with moving plate C) is used to perform interference, where the two HWPs are both set as $22.5^\circ$ to project the photons in $(\vert H\rangle+\vert V\rangle)/\sqrt2$, with one output of the beam splitter sent to PBS for coincidence. The other side (without C) is for trigger. \textbf{b.} The corresponding four-photon interference dip through adjustment of plate C.}\label{BSM}
\end{figure}

By adjusting the delay of the pink photon, we could measure the HOM dip to claim the temporal indistinguishability as shown in Fig.~\ref{HOM}{\bf b}, with the visibility of 0.9734 $\pm$ 0.0032, where the visibility is calculated as $\mathcal{V}=(\mathcal{C}_{\mathrm{max}}-\mathcal{C}_\mathrm{min})/(\mathcal{C}_\mathrm{max}+\mathcal{C}_\mathrm{min})$, and $\mathcal{C}_\mathrm{max}$($\mathcal{C}_\mathrm{min}$) corresponds to the maximum (minimum) of coincidence counts (all coincidence results do not subtract accidental coincidence counts and all error bars are estimated as standard	deviations of photon counts assuming a Poisson distribution). Similarly, for Source II in Fig.~\ref{HOM}{\bf c}, the corresponding visibility of the HOM dip has the value of 0.9593 $\pm$ 0.0045 in Fig.~\ref{HOM}{\bf d}.

The further step is to ensure the indistinguishability among the four independent photons. As shown in Fig.~\ref{BSM}, photons at the side with a movable platform C are meeting at the beam splitter for interference, whilst photons on the other side are just for trigger. By adjustment of the delay through C, the corresponding interference dip with visibility 0.8436 $\pm$ 0.0405 is reported in Fig.~\ref{BSM}{\bf b}. At this stage, an overall source of four indistinguishable independent photons just based on the identity of uncorrelated photons coming from Sources I and II is confirmed.

\section*{APPENDIX D: Calculation of photons' balanced sending probability}\label{app:D}

We introduce the method to optimize the balance of the sending probability of each single photon to the nodes. Here, for Source I, we mark $(\pi/2-\theta_1)/2$ as the angle of the HWP tuning the horizontally polarized photon (the red spin in the main text Fig.~2), and $\theta_2/2$ as the HWP tuning the vertically polarized photon (the blue spin in the main text Fig. 2). For Source II, the angles of the corresponding two HWPs are noted as $(\pi/2-\phi_1)/2$ (the purple spin in the main text Fig.~2) and $\phi_2/2$ (the orange spin in the main text Fig.~2). Thus the desired global state $\ket{\alpha H, \alpha^\prime V, \beta H,\beta^\prime V}$ is in the form of $|(\cos\theta_1 \rm R+\sin\theta_1 M){\it H},(\sin\theta_2 \rm R+\cos\theta_2 M){\it V},(\cos\phi_1 L+\sin\phi_1 M){\it H},(\sin\phi_2 L+\cos\phi_2 M){\it V}\rangle$. Then, the post-selected state can be written as (omitting the normalization)
\begin{equation}
	\begin{split}
		\ket{\Psi^{(4)}_\mathrm{PS}}=&t_1 \ket{ \rm M {\it H}, M {\it V},R{\it V},L{\it H}}+\\
		&t_2\ket{\rm M{\it V}, M{\it H},R{\it H},L{\it V}}+\\
		&t_3\ket{\rm M{\it H}, M{\it H},R{\it V},L{\it V}}+\\
		&t_4\ket{\rm M{\it V}, M{\it V},R{\it H},L{\it H}},\\
	\end{split}
\end{equation}
where
\begin{equation}
	\begin{split}
		&t_1=\sin\theta_1\cos\phi_2\sin\theta_2\cos\phi_1,\\
		&t_2=\cos\theta_2\sin\phi_1\cos\theta_1\sin\phi_2,\\
		&t_3=\sin\theta_1\sin\phi_1\sin\theta_2\sin\phi_2\\
		&t_4=\cos\theta_2\cos\phi_2\cos\theta_1\cos\phi_1.
	\end{split}
\end{equation}
The state above can be finally recast as (omitting the normalization)
\begin{equation}
	\begin{split}
		\ket{\Psi^{(4)}_\mathrm{PS}}=&\ket{\Psi_\mathrm{M},t_1 \rm R{\it V},\rm L{\it H}+t_2\rm  R{\it H},\rm L{\it V}}+\\
		&\ket{\Phi^+_\mathrm{M},t_3 R{\it V},L{\it V}+t_4 \rm R{\it H},\rm L{\it H}}-\\
		&\ket{\Phi^-_\mathrm{M},t_3 \rm R{\it V},\rm L{\it V}-t_4 \rm R{\it H},\rm L{\it H}},\\
	\end{split}
\end{equation}
with $\ket{\Psi_\mathrm{M}}=\ket{ \rm M {\it H}, M {\it V}}$ and $\ket{\Phi^\pm_\mathrm{M}}=(\ket{ \rm M {\it H}, M {\it H}}\pm\ket{ \rm M {\it V}, M {\it V}})/\sqrt2$.

Now, if each photon is sent to the nodes with a balanced probability, which means the angles of $\theta_1$, $\theta_2$, $\phi_1$ and $\phi_2$ all equal to $\pi/4$, the spatial wavefunctions of photons from the same source are completely overlapped, i.e., the degree of indistinguishability of photons are maximal, and the final state is obtained with a maximal probability. Otherwise, the spatial wavefunctions are partially overlapped which leads to get the final state with a lower probability.

\section*{appendix e: Detailed explanation of the localized product-state measurement}

Here we describe the details of LPSM in the mode M.
Firstly, we describe the transformations of three states $\ket{\Psi_\mathrm{M}}$, $\ket{\Phi_\mathrm{M}^+}$ and $\ket{\Phi_\mathrm{M}^-}$ due to the utilization of two 50:50 beam splitters (BS).
Writing the operation matrix of the BS as
$\frac{1}{\sqrt{2}}\left(\begin{smallmatrix} 1 & i \\  i & 1 \end{smallmatrix} \right),$ which depicts the mapping between two outputs and two inputs of the BS, the state $\ket{\Psi_\mathrm{M}}=\ket{H, V}$ becomes $\ket{H, i V}$ after the first BS. When impinging the second BS with the labels of two output paths, the state is transformed to be $(\ket{H_1, i^2 V_2}+\ket{i H_2, i V_1}+\ket{H_1, i V_1}+\ket{i H_2, i^2 V_2})/2$ which could be simplified as $(\ket{H_1, V_2}+\ket{V_1, H_2}-i \ket{H_1, V_1}+i \ket{H_2, V_2})/2$.
Since we perform the product-state measurement between the path 1 and path 2 and the signals of two corresponding detectors are handled by the coincidence device, the cases in which two photons are located in the same path, i.e., the states $\ket{H_1, V_1}$ and $\ket{H_2, V_2}$, are discarded in this post-selection measurement. In other words, only the coincidence count coming from the signals of two detectors placed on path 1 and path 2 is valid. Thus, the measurement-induced state on node M could be written as $\ket{\Psi'_\mathrm{M}}=(\ket{H_1, V_2}+\ket{V_1, H_2})/\sqrt2$. Under the same framework, the states $\ket{\Phi_\mathrm{M}^\pm}$ are transformed to the measurement-induced states $\ket{\Phi_\mathrm{M}'^{\pm}}=(\ket{H_1, H_2}\pm \ket{V_1, V_2})/\sqrt2$.

Based on the above description, the effort to distinguish the three states $\ket{\Psi_\mathrm{M}}$ and $\ket{\Phi_\mathrm{M}^\pm}$ is transformed to distinguish the measurement-induced states $\ket{\Psi'_\mathrm{M}}$ and $\ket{\Phi_\mathrm{M}'^\pm}$. Here, we choose an appropriate LPSM on M node, in which two of three states $\ket{\Psi'_\mathrm{M}}$ and $\ket{\Phi_\mathrm{M}'^\pm}$ vanish while the other one is nonzero, as explicitly described in the main text.
In particular: (i) measuring the product state $\ket{H_1} \otimes \ket{V_2}$ (or $\ket{V_1} \otimes \ket{ H_2}$) on node M, the outputs of $\ket{\Phi_\mathrm{M}'^{\mp}}$ are zero while the outcomes of $\ket{\Psi'_\mathrm{M}}$ would exhibit coincidence at the two final detectors in M; (ii) projecting on $\ket{r_1} \otimes \ket{l_2}$ (or $\ket{l_1} \otimes \ket{r_2}$), where $\ket{r}=(\ket{H}+i\ket{V})/\sqrt2$ and $\ket{l}=(\ket{H}-i\ket{V})/\sqrt2$, we pick up $\ket{\Phi_\mathrm{M}'^{+}}$ since $\ket{\Phi_\mathrm{M}'^{-}}$ and $\ket{\Psi'_\mathrm{M}}$ vanish; (iii) the state $\ket{\Phi_\mathrm{M}'^{-}}$ is filtered out by measuring the product state $\ket{d_1} \otimes \ket{c_2}$ (or $\ket{c_1} \otimes \ket{d_2}$), where $\ket{d}=(\ket{H}+\ket{V})/\sqrt2$ and $\ket{c}=(\ket{H}-\ket{V})/\sqrt2$.

\end{document}